\documentclass[aps,prd,nofootinbib,twocolumn,floatfix,nobalancelastpage,
letterpaper,showpacs]{revtex4}
\usepackage{graphicx}\usepackage{natbib}\usepackage{amsmath}
\begin{document}

\title{Gamma-ray signatures of annihilation to charged leptons in dark matter substructure}

\author{Matthew D. Kistler}
%\email{kistler@mps.ohio-state.edu}
%\affiliation{Department of Physics, Ohio State University, Columbus, Ohio 43210}
\affiliation{Center for Cosmology and Astro-Particle Physics and Department of Physics, Ohio State University, Columbus, Ohio 43210}

\author{Jennifer M. Siegal-Gaskins}
%\email{jsg@mps.ohio-state.edu}
\affiliation{Center for Cosmology and Astro-Particle Physics and Department of Physics, Ohio State University, Columbus, Ohio 43210}

\date{March 17, 2010}

\begin{abstract}
Due to their higher concentrations and small internal velocities, Milky Way subhalos can be at least as important as the smooth halo in accounting for the GeV positron excess via dark matter annihilation.  After showing how this can be achieved in various scenarios, including in Sommerfeld models, we demonstrate that, in this case, the diffuse inverse-Compton emission resulting from electrons and positrons produced in substructure leads to a nearly-isotropic signal close to the level of the isotropic GeV gamma-ray background seen by Fermi.  Moreover, we show that HESS cosmic-ray electron measurements can be used to constrain multi-TeV internal bremsstrahlung gamma rays arising from annihilation to charged leptons.
\end{abstract}

% 95.35.+d     Dark matter
% 95.85.Ry     Neutrino, muon, pion, and other elementary particles; cosmic rays
% 98.70.Sa		 Cosmic rays
% 98.70.-f 	   Unidentified sources of radiation outside the Solar System
%,showpacs
\maketitle

%--------------------------------------------------------------------%
\section{Introduction}
The possibility of using dark matter~\cite{Zwicky:1933gu} to provide energetic positrons and electrons~\cite{Silk:1984zy,Tylka:1989xj,Rudaz:1987ry,Eichler:1989br,Kamionkowski:1990ty} to explain the GeV positron excess~\cite{HEAT} seen by PAMELA~\cite{Adriani:2008zr}, the ``ATIC bump''~\cite{Chang:2008zz,Torii:2008xu}, and the less-anomalous $e^- + e^+$ spectra measured by the Fermi Gamma-ray Space Telescope (hereafter, Fermi)~\cite{Fermi:2009zk} and HESS~\cite{Aharonian:2008aaa,Aharonian:2009ah} (in lieu of a nearby pulsar~\cite{Yuksel:2008rf,Hooper:2008kg,Profumo:2008ms}) has sparked considerable interest (e.g., \cite{ArkaniHamed:2008qn,Pospelov:2008jd,Lattanzi:2008qa,Cirelli:2008pk,Nomura:2008ru,Feldman:2008xs,Barger:2008su,Harnik:2008uu,Kohri:2009bd,Rothstein:2009pm,Ibe:2009dx,Grasso:2009ma}).  Generating the required $e^\pm$ flux through annihilations in the smooth component of the Milky Way's dark matter halo requires both a much larger annihilation cross section than might be expected for a thermal relic \cite{Jungman:1995df} and a large branching ratio to charged leptons.  A variety of constraints (e.g., \cite{Cholis:2008wq,Zhang:2008tb,Borriello:2009fa,Pato:2009fn,Cirelli:2009vg,Zhang:2009pr,Pohl:2009qt}) already apply to a smooth halo explanation, and will likely tighten with new data from Fermi~\cite{Atwood:2009ez}.

We examine the observational consequences of annihilations occurring within the dark matter substructure of the Milky Way.  Substructure differs from the smooth halo in its spatial distribution, which is less centrally-concentrated~\cite{Diemand:2008in,Springel:2008cc}, and in its characteristic velocity dispersions, which are colder.  We focus on the case of enhanced annihilation to charged leptons in ``Sommerfeld'' models, in which the annihilation cross section increases with decreasing relative velocity, and consider scenarios with a velocity-independent cross section.  Annihilations to charged particles necessarily produce internal bremsstrahlung (IB) gamma rays \cite{Beacom:2004pe,Bergstrom:2008ag,Bell:2008vx,Essig:2009jx,Pieri:2009zi}.  High-energy electrons and positrons will also produce gamma rays through various energy-loss processes.  Beyond a few tens of kiloparsecs from the Galactic Center, the dominant loss process for electrons is the inverse-Compton (IC) upscattering of CMB photons.

We calculate the expected high-latitude gamma-ray emission from IB and IC resulting from dark matter annihilation in substructure throughout the Milky Way.  In particular, rather than considering only the serendipitous presence of a single, nearby, massive dark matter clump~\cite{Hooper:2008kv,Brun:2009aj}, we account for the collective emission from the entire subhalo population.  Noting that electron- and gamma-ray-induced showers are difficult to distinguish in air Cherenkov telescopes, we discuss how constraints on TeV gamma-ray fluxes can be obtained from the cosmic-ray electron measurements made by HESS to limit annihilations to lepton pairs via their IB emission.

%--------------------------------------------------------------------%
\section{Annihilation in Substructure}
To describe their structural properties, we assume each individual dark matter subhalo to be described by a NFW density profile~\cite{Navarro:1994hi},
\begin{equation}
  \rho_{\rm sub}=\frac{\rho_s}{r/r_s \left(1+r/r_s\right)^2} \,,
\label{NFW}
\end{equation}
where $r_{s}$ is a scale radius and $\rho_s$ a characteristic density.  The differential luminosity (photons or particles per energy per time) $L$ of a subhalo, for an annihilation cross section $(\sigma v)_0$ that is independent of velocity, is
\begin{equation}
\label{eq:lsub}
  L = K \int \,dV_{\rm sub}\, \rho^{2}_{\rm sub} \propto \rho_s^2\,r_s^3
    \propto M\,\frac{c^3}{f^2(c)},
\end{equation}
where $M$ is the subhalo mass, $c=r_{\rm vir}/r_s$ is the concentration, $f(c)=\ln(1+c)-c/(1+c)$, and the particle physics-dependence of the annihilation rate is isolated in
\begin{equation}
  K=\frac{(\sigma v)_0}{2m_{\rm DM}^{2}} \frac{dN}{dE},
\end{equation}
with $m_{\rm DM}$ the dark matter particle mass and $dN/dE$ the particle spectrum produced per annihilation.

Numerical simulations find a relation between concentration and mass for subhalos that varies as a function of distance from the Galactic Center~\cite{Diemand:2008in,Springel:2008cc}.  This is a natural consequence of tidal stripping, which more effectively removes mass from the outer regions of the subhalos while leaving the core relatively unscathed.  Thus, for a given subhalo mass, subhalos located nearer to the Galactic Center will be more luminous than those at large radii.  We adopt the modified Bullock et al.~\cite{Bullock:1999he} relation for low-mass halos, with radial dependence, from Ref.~\cite{Kuhlen:2008aw}
\begin{equation}
\label{eq:cmrsub}
  c_{\rm sub}(M,r) = 18\, \left(\frac{M}{10^8\,M_\odot}\right)^{-0.06}\,
                \left(\frac{r}{r_{\rm field}}\right)^{-0.286},  
\end{equation}
where $r_{\rm field} = 402$~kpc is the radius where equal mass subhalos and field halos have the same concentration (see also \cite{Pieri:2007ir}).  Using this relation with Eq.~(\ref{eq:lsub}), we can approximate the differential luminosity of a subhalo of mass $M$ at a radius $r$ from the Galactic Center by
\begin{equation}
\label{eq:lsubmr}
  L(M,r)=K\,\mathcal{L}(M)\,\left(\frac{r}{r_{\rm field}}\right)^{-0.7},
\end{equation}
where we have defined
\begin{equation}
\label{eq:lofm}
  \mathcal{L}(M)=\int dV_{\rm sub}\, \rho^{2}_{\rm sub}
       \simeq \mathcal{L}_{0} \,  \left(\frac{M}{M_{\rm 0}}\right)^{0.87}
\end{equation}
to describe the dependence of the annihilation rate on the structural properties of the subhalo.  We note that the dependences on $M$ and $r$ are not formally separable, but are weak enough that Eq.~(\ref{eq:lsubmr}) gives a reasonable approximation.  The mass of Canes Venatici I \cite{Simon:2007dq}, assuming a NFW density profile and concentration $c=19.5$, is used to normalize $\mathcal{L}_{0}$ and $M_{0}$.

%
%%%%%%%%%%%%%%%%%%%%%%%%%
\begin{figure}[b!]
\includegraphics[width=3.25in,clip=true]{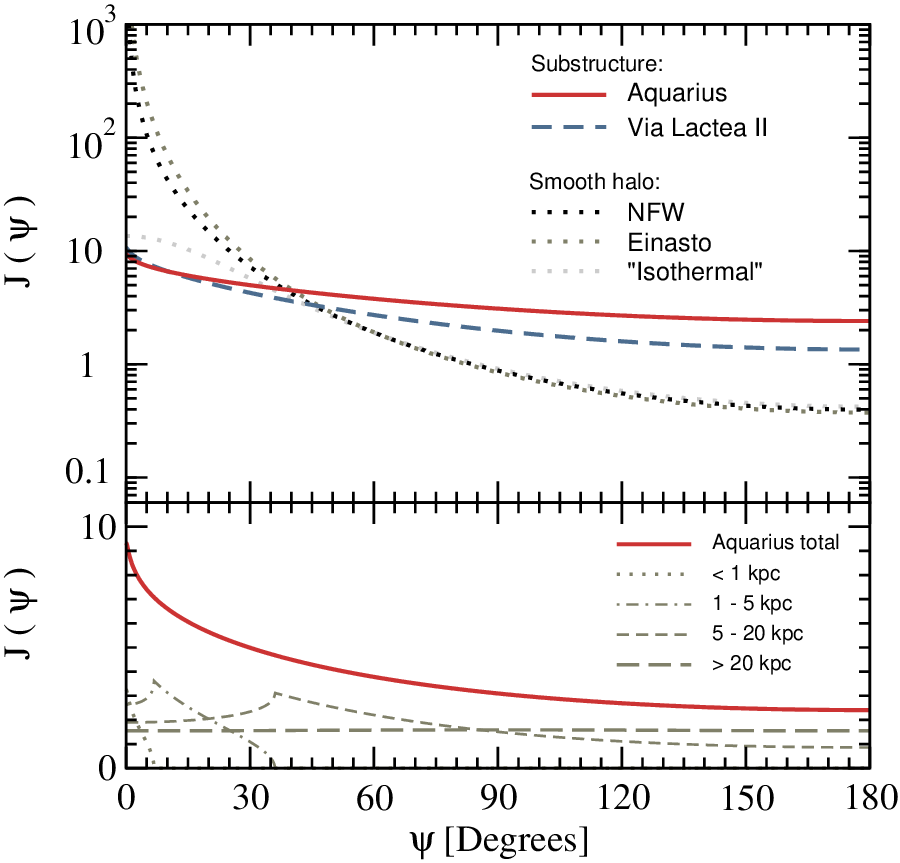}
\caption{Angular distribution of the emission from dark matter annihilation (${\cal J}(\psi)$ from Eq.~(\ref{Jeq})).  {\it Top panel}:  Galactic substructure assuming radial distributions from Aquarius ({\it solid}) and Via Lactea II ({\it dashed}), compared to the smooth halo ({\it dotted}) assuming NFW, Einasto ($\alpha=0.17$, $r_{-2}=20$~kpc), and cored isothermal ($r_{\rm core}=5$~kpc) density profiles (as labeled).
{\it Bottom panel}: The substructure emission assuming the Einasto radial distribution with Aquarius parameters, broken down into contributions from radial shells surrounding the Galactic Center (with distances as labeled).  Peaks are due to shell boundaries.
\label{Jpsi}}
\end{figure}
%%%%%%%%%%%%%%%%%%%%%%%%%
%

We assume a power-law mass function for the subhalos~\cite{Diemand:2008in,Springel:2008cc}, $dN/dM \propto M^{-\alpha}$ with $\alpha = 1.9$ and extrapolate this relation to a minimum subhalo mass $M_{\rm min}=10^{-6}$ $M_{\odot}$.  Noting that
$(dN/d\mathcal{L})=(dN/dM)(dM/d\mathcal{L})$, integration over the subhalo population yields
\begin{equation}
\label{eq:lsubs}
  \mathcal{L}_{_{\rm subs}}=\int_{\mathcal{L}_{_{\rm min}}}^{^{\mathcal{L}_{\rm max}}} \!\!\!
    d\mathcal{L}\, \mathcal{L}\, \frac{dN}{d\mathcal{L}}\,,
\end{equation}
which contains the dependence of the annihilation rate on the structural properties and mass function of the subhalos, and is independent of position in the Galaxy.

We model the subhalo number density (i.e., number of subhalos per volume) at a radius $r$ from the Galactic Center with an Einasto profile~\cite{Einasto}
\begin{equation}
  n_{_{\rm subs}}(r) \propto \exp\left\{-\frac{2}{\alpha_{_{\rm subs}}}
    \left[ \left( \frac{r}{r_{-2}} \right)^{\alpha_{_{\rm subs}}} -1 \right]  \right\},
\label{nein}
\end{equation}
with $\alpha_{_{\rm subs}}=0.68$ and $r_{-2}=199$~kpc, as found by the Aquarius Project~\cite{Springel:2008cc}.  We assume a NFW profile for the smooth halo with $r_s = 20$~kpc, $r_{\rm vir} = 255$~kpc, and $M_{\rm vir} = 1.9 \times 10^{12}\,M_\odot$, and normalize the subhalo distribution such that a fraction $f_{\rm subs}=0.15$ of $M_{\rm vir}$ is bound in substructure.  
For comparison, we also consider a number density distribution as in Via Lactea II, $n_{_{\rm VL-II}}(r) \propto (1 + r/R_s)^{-2}$~\cite{Kuhlen:2008aw}, with $R_s \approx 20$~kpc.

The intensity ($I$) of gamma-ray emission resulting from annihilation in substructure at an angle $\psi$ from the Galactic Center is
\begin{equation}
  I(\psi) = \frac{K}{4\pi} r_{\odot} \rho_\odot^2 \cal{J}(\psi),
\label{Ieq}
\end{equation}
in which $r_\odot=8.5$~kpc and ${\cal J}(\psi)$ is given by the line-of-sight integral
\begin{equation}
 {\cal J}(\psi) = \frac{\mathcal{L}_{_{\rm subs}}}{r_\odot \rho_\odot^2}
     \int_{los} ds \, n_{_{\rm subs}}(r)\, \left(\frac{r}{r_{\rm field}}\right)^{-0.7},
\label{Jeq}
\end{equation}
where $r=r(s,\psi)$.  This is shown in Fig.~\ref{Jpsi} for the Aquarius and VL-II subhalo radial distributions, along with the analogous term for three possible smooth halo density profiles.  Here, we have chosen $\mathcal{L}_{_{\rm subs}}$ such that the annihilation rate per volume is matched to the smooth halo at $r_\odot$.  Both subhalo models produce a nearly-isotropic angular signal, much less strongly-peaked towards the Galactic Center than that of the smooth halo.  In the following, we adopt the Aquarius distribution from Eq.~(\ref{nein}).

%--------------------------------------------------------------------%
\section{Sommerfeld enhancement in substructure and the smooth halo}
A challenge in attributing the positron excess to dark matter annihilation is the need for a much larger annihilation cross section than expected for a thermal relic.  One way to accomplish this is by introducing a scalar or vector boson that mediates an intermediate-range force between dark matter particles and can dramatically enhance the cross section for annihilation at low relative velocities (see Sommerfeld~\cite{Sommerfeld(1931)} for electromagnetic scattering).  For dark matter, this can be through Standard Model gauge bosons (e.g., \cite{Hisano:2004ds,Lattanzi:2008qa}) or a new mediator particle ($\phi$)~\cite{ArkaniHamed:2008qn,Pospelov:2008jd}.

We examine the possibility that annihilations in substructure, rather than in the smooth halo, are the dominant contributor to the measured local lepton flux and the astrophysical consequences that result.  In this Section, we first show how this can arise for Sommerfeld models, then proceed to properties valid in general.  The Sommerfeld enhancement can be expressed by the factor $S(v)=(\sigma v)/(\sigma v)_{0}$, where $v$ is the relative velocity of the dark matter particles and $(\sigma v)_{0}$ is the cross section in the absence of the additional force.  These models can be parametrized by the coupling of the dark matter to the mediator ($\alpha_{S}$) and the ratio of the dark matter and mediator masses, $m_{\rm DM}/m_{\phi}$.  For small velocities~\cite{ArkaniHamed:2008qn},
\begin{equation}
  S \sim \frac{\pi\alpha_{S}}{v}.
\end{equation}
The enhancement saturates at the velocity $v_{\rm min}$ at which the de Broglie wavelength of the dark matter particles becomes comparable to the range of the force, i.e. $1/(m_{\rm DM}\,v) \sim 1/m_{\phi}$.  The maximum enhancement is
\begin{equation}
  S_{\rm max} \sim \pi\alpha_{S} \frac{m_{\rm DM}}{m_{\phi}}.
\end{equation}
Resonances corresponding to bound states can result in significantly larger enhancements in such models, when $m_{\rm DM}/m_{\phi}\simeq 2 n^2/\alpha_{S}$, for integer $n$~\cite{Lattanzi:2008qa}.  We account for this by solving the nonrelativistic Schr{\"o}dinger equation as in \cite{ArkaniHamed:2008qn,Lattanzi:2008qa,Robertson:2009bh}.  For simplicity, we consider $\alpha_{S} = 10^{-2}$, and note that this choice qualitatively demonstrates the features of these models.

%
%%%%%%%%%%%%%%%%%%%%%%%%%
\begin{figure}[b!]
\includegraphics[width=3.25in,clip=true]{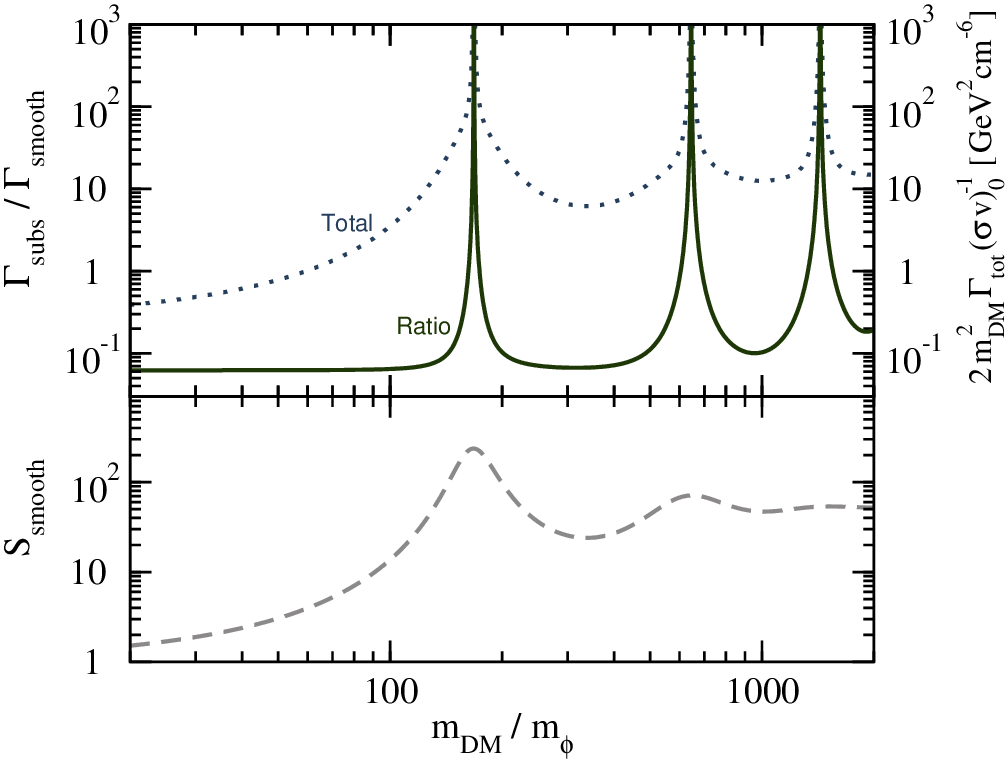}
\caption{Subhalo and smooth halo contributions to the local annihilation rate as a function of $m_{\rm DM}/m_{\phi}$ in models with resonant behavior, for $\alpha_{\rm S} = 10^{-2}$.  {\it Top panel:} Ratio of the annihilation rate in substructure to the smooth halo ({\it solid, left axis}) and total annihilation rate ({\it dotted, right axis}).  {\it Bottom panel:} Local enhancement factor in the smooth halo.
\label{fig:localratiores}}
\end{figure}
%%%%%%%%%%%%%%%%%%%%%%%%%
%

We also consider a more general parametrization of $1/v$ models without resonant behavior, as in \cite{Kuhlen:2009is}:
\begin{equation}
  S(v)=S_{\odot}\, \frac{v_{\odot}}{v+v_{\rm min}},
\end{equation}
where $v_{\odot}$ and $S_{\odot}$ are the relative velocity of smooth halo particles and enhancement factor at $r_{\odot}$.  Without allowing for resonant behavior, large values of $m_{\rm DM}/m_{\phi}$ are required to produce large enhancements.

\begin{figure}[t!]
\includegraphics[width=3.25in,clip=true]{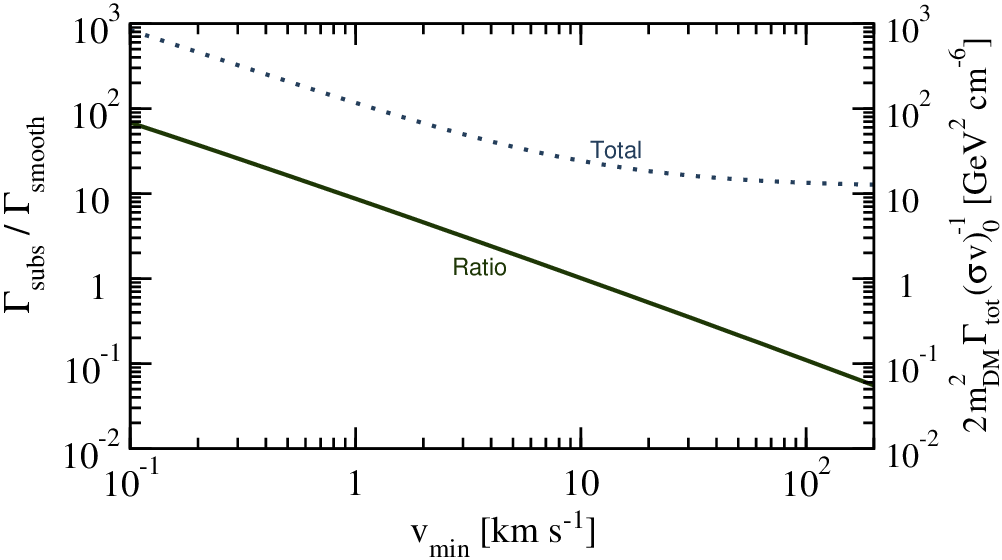}
\caption{Same as top panel of Fig.~\ref{fig:localratiores}, but as a function of $v_{\rm min}$ for a Sommerfeld model without resonant behavior.  The total rate ({\it dotted line, right axis}) is calculated for $S_{\odot}=50$ (both the smooth halo and subhalo rates scale linearly with $S_{\odot}$).
\label{fig:localratiogen}}
\end{figure}

In the smooth halo, the local annihilation rate per volume, $\Gamma_{_{\rm smooth}}$, depends on the dark matter density and velocity dispersion at the solar circle.  We normalize the 1-D velocity dispersion to the local rotation curve~\cite{Reid:2009nj}, $\sigma_{\rm 1D}\simeq v_{\rm circ}/\sqrt{2}\approx 180$~km~s$^{-1}$, and take $\sigma_{\rm 1D} \sim v$.  Then
\begin{equation}
\label{eq:ratesm}
  \Gamma_{_{\rm smooth}} =  \frac{\langle \sigma v \rangle_{0}\, S_{\odot}}{2 m_{\rm DM}^{2}}\,
                     \rho^{2}_{\odot}.
\end{equation}
As subhalos are dynamically colder than the smooth halo, their smaller internal velocity dispersions can lead to preferential enhancement of the annihilation rate.  We adopt a simple relation between subhalo mass and velocity dispersion, $\sigma_{v} \sim M_{\rm sub}^{1/3}$ (see also \cite{Bovy:2009zs}), and assign an enhancement factor $S[v(M_{\rm sub})]$ to each subhalo of a given mass by approximating the relative velocity of the particles throughout the subhalo by $\sigma_{v} \sim v$.  (The $r$ dependence of $c_{\rm sub}$ in Eq.~(\ref{eq:cmrsub}) leads to a mild variation that we neglect here.)  We normalize the $\sigma_{v}(M_{\rm sub})$ relation using again the mass and velocity dispersion of Canes Venatici I \cite{Simon:2007dq}.  For Sommerfeld models, Eq.~(\ref{eq:lofm}) becomes
\begin{equation}
\label{eq:slofm}
  \mathcal{L}(M)=\mathcal{L}_{0} \, \left(\frac{M}{M_{\rm 0}}\right)^{0.87}S[\sigma_{\rm v}(M)]\,.
\end{equation}
The value of $\mathcal{L}_{_{\rm subs}}$ is then determined by Eq.~(\ref{eq:lsubs}) using $\mathcal{L}(M)$ as defined in Eq.~(\ref{eq:slofm}).  The local annihilation rate per unit volume from subhalos is
\begin{equation}
\Gamma_{_{\rm subs}} =  \frac{\langle \sigma v \rangle_{0}}{2 m_{\rm DM}^{2}}\,
  \mathcal{L}_{_{\rm subs}}\, n_{_{\rm subs}}(r_{\odot})\, \left(\frac{r_{\odot}}{r_{\rm field}}\right)^{-0.7}.
\end{equation}
The large local number density of subhalos ($n_{_{\rm subs}}(r_{\odot}) \sim 2 \times 10^{8}$ kpc$^{-3}$) allows us to neglect the discreteness of subhalos as sources and use the annihilation rate averaged over the subhalo population via $\mathcal{L}_{_{\rm subs}}$.  Fig.~\ref{fig:localratiores} compares the local annihilation rates in the smooth halo and in substructure for a Sommerfeld model with $\alpha_{S} = 10^{-2}$.  For a significant region of the $m_{\rm DM}/m_{\phi}$ parameter space, particularly near resonances, the subhalo rate vastly exceeds the smooth halo rate ({\it top panel, solid line}).  In Fig.~\ref{fig:localratiogen}, the local smooth halo and substructure annihilation rates for a model without resonant behavior are compared.  The two contributions are comparable at $v_{\rm min} \sim 10$ km/s, with subhalos dominating at lower $v_{\rm min}$.

At high subhalo masses ($\gtrsim 10^7\,M_\odot$) this approach breaks down due to the low number density of these subhalos, so that the presence of a nearby clump would have to be accounted for (although having such within 1~kpc is unlikely~\cite{Diemand:2008in}).  Including subhalos of all masses in the calculation of the local flux results in an overestimate of $\mathcal{L}_{_{\rm subs}}$ of not more than 25\% relative to assuming a maximum mass of $10^7\,M_\odot$ without accounting for a Sommerfeld enhancement, and smaller in models with lower $v_{\rm min}$, a minor effect that we neglect.  This can be understood from Eq.~(\ref{eq:lsubs}): in absence of a Sommerfeld enhancement, the total luminosity roughly scales $\propto \int dM\, M^{0.87}\, dN/dM \propto M_{\rm min}^{-0.03}$, so that each decade of mass contributes nearly equally.  (This has a much weaker dependence on the choice of lower cutoff than the $L_{\rm tot}\propto M_{\rm min}^{-0.226}$ of Ref.~\cite{Springel:2008by}.)

%
%%%%%%%%%%%%%%%%%%%%%%%%%
%%%%%%%%%%%%%%%%%%%%%%%%%
%

The choices we have made in defining our substructure model are conservative: using a steeper mass function (e.g., $\alpha=2$), a smaller minimum subhalo mass $M_{\rm min}$~(e.g., \cite{Green:2005fa,Profumo:2006bv,Bringmann:2009vf}) a larger mass fraction in substructure $f_{\rm sub}$, a higher normalization of the $c(M)$ relation (as in Aquarius), or considering substructures-within-substructure~\cite{Diemand:2008in,Kuhlen:2008aw,Springel:2008by} would all increase the subhalo annihilation rate relative to the smooth halo.  The dependence of the ratio of the local annihilation rates in substructure and the smooth halo $\Gamma_{\rm subs}/\Gamma_{\rm smooth}$ on $\alpha$, $f_{\rm sub}$, and $M_{\rm min}$ is explored in Table~\ref{tab:subparams}, assuming a velocity-independent cross section (i.e., no Sommerfeld enhancement).  For $\alpha = 1.9$ (2.0) the mass function is normalized so that 15\% (50\%) (e.g., \cite{Kuhlen:2008aw}) of the host halo mass is in subhalos of $10^{-6}$ to $10^{10}$ M$_{\odot}$.

The range of $\Gamma_{\rm subs}/\Gamma_{\rm smooth}$ in Table~\ref{tab:subparams} indicates that even in the absence of an enhancement arising primarily from Sommerfeld effects substructure can contribute significantly to the local annihilation rate, and hence no more than a modest enhancement in the substructure rate is typically required for this component to be locally dominant (see also \cite{Yuan:2009bb,Pieri:2009je}).  Current CMB constraints~\cite{Slatyer:2009yq,Galli:2009zc} restrict the saturation cross section in Sommerfeld models to within a factor of a few of the value required in order to explain the PAMELA and Fermi data by annihilation in the smooth halo, consequently only models in which the subhalo contribution is no more than a factor of a few smaller than that of the smooth halo without Sommerfeld enhancement could produce sufficient cosmic-ray fluxes without exceeding the allowed saturation cross section.  Clearly, this condition can be satisfied for plausible subhalo parameters, which is valid even for dark matter models with other means of obtaining an enhanced annihilation cross section.  We proceed assuming that annihilation to charged leptons in substructure dominates over the smooth halo and accounts for 100\% of the anomalous fluxes.  This condition is sufficient for the following purposes, which can be rescaled linearly with the local annihilation rate in substructure as desired.

\begin{table}[t!]
\caption{\label{tab:subparams}The ratio of local annihilation rates in substructure and the smooth halo $\Gamma_{\rm subs}/\Gamma_{\rm smooth}$ and the fraction of the halo mass bound in substructure $f_{\rm sub}$ for various choices of mass function slope $\alpha$ and minimum subhalo mass $M_{\rm min}$, in the absence of Sommerfeld enhancement.  The maximum subhalo mass is $10^{10}$ M$_{\odot}$ in all cases.}
\begin{ruledtabular}
\begin{tabular}{cccc}
$\alpha$ & $M_{\rm min}$ (M$_{\odot}$) & $f_{\rm sub}$ & $\Gamma_{\rm subs}/\Gamma_{\rm smooth}$\\
\hline
1.9 & $10^{-4}$ & 0.148 & 0.0366\\
1.9 & $10^{-6}$ & 0.150 & 0.0452\\
1.9 & $10^{-12}$ & 0.153 & 0.0815\\
2.0 & $10^{-4}$ & 0.442 & 0.300\\
2.0 & $10^{-6}$ & 0.500 & 0.541\\
2.0 & $10^{-12}$ & 0.672 & 3.48\\
\end{tabular}
\end{ruledtabular}
\end{table}

%--------------------------------------------------------------------%
\section{Internal Bremsstrahlung}
For pure leptonic final states, the only gamma-ray emission directly resulting from annihilations in substructure is internal bremsstrahlung (IB), e.g., $\chi \chi \rightarrow \ell^+ \ell^- \gamma$.  We consider a few representative annihilation channels: direct annihilation into $2\,\mu$ and $2\,\tau$, and annihilation to $4\,\tau$ through a new particle $\phi$ ($\chi\chi \rightarrow \phi\phi$ and each $\phi \rightarrow 2\tau$), with cross sections required to explain the combined PAMELA/Fermi data~\cite{Bergstrom:2009fa,Meade:2009iu}.  We calculate the IB spectra for the two lepton cases as in Ref.~\cite{Bergstrom:2008ag},
\begin{equation}
  \frac{dN_{\rm IB}}{dE} = \frac{1}{E} \,\frac{\alpha_{\rm em}}{\pi}\, \left(1+x^2\right)
                   \ln\left( \frac{4 m_{\rm DM}^2 x}{m_{\mu,\, \tau}^2} \right) \,,
\end{equation}
where $x=1-E/m_{\rm DM}$ and $\alpha_{\rm em}\simeq 1/137$.  For $m_{\rm DM} \gg m_{\mu,\, \tau},\,E$, this has the behavior $dN_{\rm IB}/dE\propto E^{-1}$~\cite{Beacom:2004pe}.  We similarly calculate the four lepton case, as detailed in Ref.~\cite{Mardon:2009rc}.  For the $2\,\mu$ channel, we omit the gamma-ray contribution from muon decay, $\mu \rightarrow e\,\nu_{e}\,\nu_{\mu}\,\gamma$, which is negligible for the scenario considered here.

%%%%%%%%%%%%%%%%%%%%%%%%%
\begin{figure*}[t!]
\includegraphics[width=7in,clip=true]{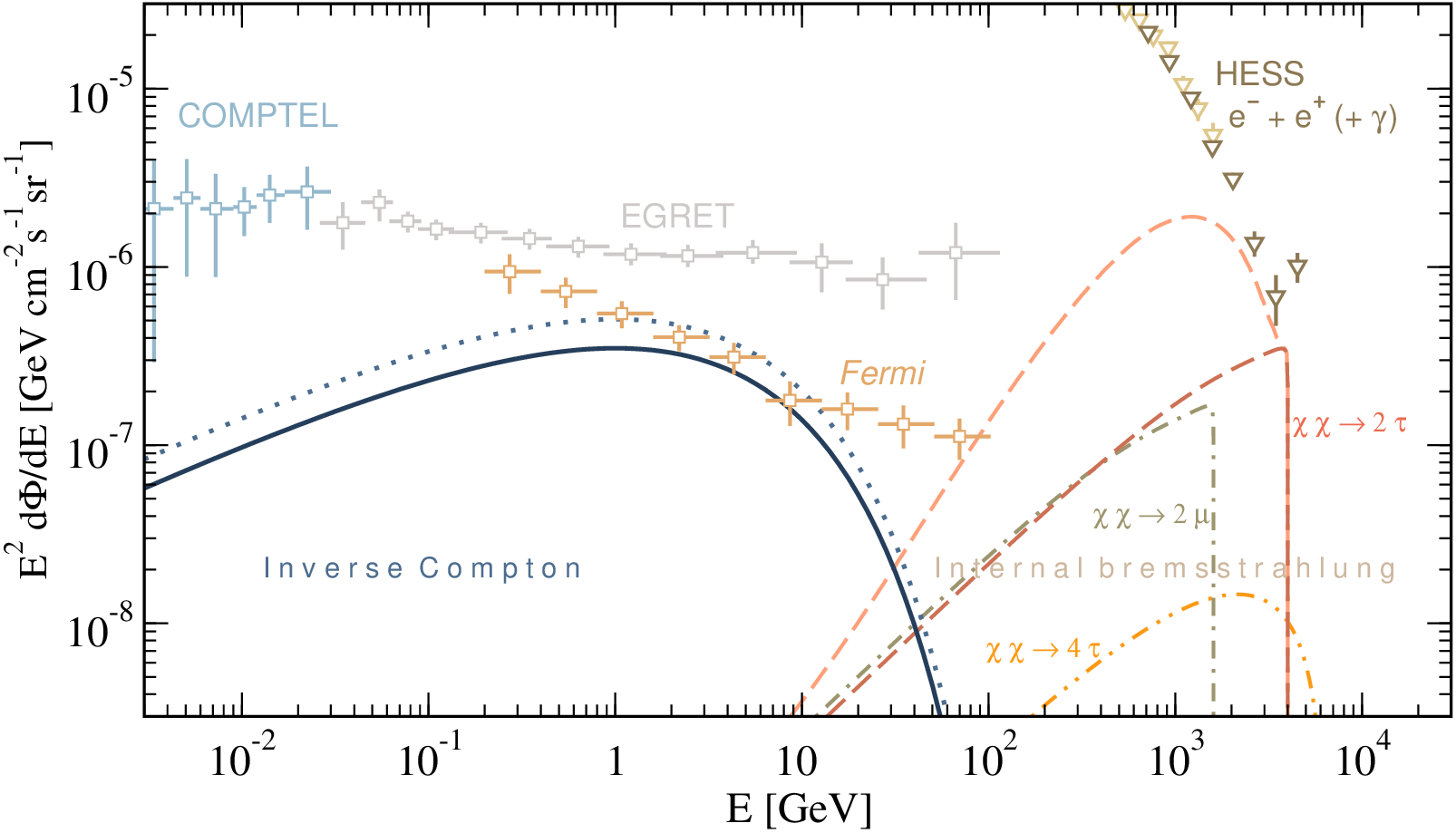}
\caption{Isotropic gamma-ray signals resulting from dark matter annihilations in substructure (assuming an Aquarius number density profile).  {\it Left side:} Inverse-Compton gamma-ray emission of the final state electron/positron population from annihilations at distances $> 20$~kpc from the Galactic Center ({\it solid line}) and including all radii ({\it dotted}).  This can be compared to COMPTEL~\cite{Weidenspointner}, EGRET~\cite{Sreekumar:1997un}, and Fermi~\cite{collaboration:2010nz} diffuse gamma-ray data.
{\it Right side:} Internal bremsstrahlung associated with the birth of charged leptons is shown for annihilation to two muons ($m_{\rm DM} = 1.6$~TeV; {\it dot-dashed}), two taus ($m_{\rm DM} = 4$~TeV; {\it dark dashed}), and two $\tau^\pm$ pairs ($m_{\rm DM} = 8$~TeV; {\it double-dot dashed}).  For the two tau case, we show the effect of including tau decays ({\it light dashed}).  Cosmic-ray $e^- + e^+$ measurements from HESS ({\it triangles})~\cite{Aharonian:2008aaa,Aharonian:2009ah} act as upper limits on an isotropic gamma-ray flux (see text).
\label{pbr}}
\end{figure*}
%%%%%%%%%%%%%%%%%%%%%%%%%

The IB gamma-ray intensity from annihilation in substructure at $\psi=180^{\circ}$ (the minimum of the dark matter signal) is shown in Fig.~\ref{pbr} for the above cases, along with gamma rays resulting directly from pionic tau decays in the $2\,\tau$ scenario (using DarkSUSY~\cite{Gondolo:2004sc}).  Considering a smooth halo model would result in signals smaller by a factor of $\sim 2-3$ for all profiles at large angles.  Although we do not otherwise consider them, models based on decays in the smooth halo (e.g., \cite{Chen:2008yi,Yin:2008bs,Arvanitaki:2008hq,Chen:2009ew}) would give signals comparable to those shown in Fig.~\ref{pbr}.

Directly measuring such a diffuse gamma-ray flux at TeV energies is presently challenging, in part due to the effective area of Fermi saturating with energy~\cite{Atwood:2009ez}.  While ground-based air Cherenkov telescopes do not have this problem, the electromagnetic showers that they observe are quite similar for TeV electrons and gamma rays, making them difficult to separate.  Based on observations of fields far from the Galactic plane, HESS has recently reported measurements of the $e^- + e^+$ spectrum into the TeV regime~\cite{Aharonian:2008aaa,Aharonian:2009ah}.  In principle, a nearly-isotropic $\sim\,$TeV gamma-ray flux from dark matter annihilation could result in an {\it apparent} feature in this spectrum.  As noted in Refs.~\cite{Aharonian:2008aaa,Aharonian:2009ah}, there should be little contribution from extragalactic TeV gamma rays.

In Fig.~\ref{pbr}, we show the HESS $e^- + e^+$ spectrum, which can be regarded as a conservative upper limit on isotropic TeV gamma rays.  The maximum gamma-ray fraction of this measurement is likely $\lesssim 10$\% (although systematic uncertainties could result in as much as $\sim 50$\%)~\cite{Aharonian:2008aaa}.  It is likely that a dedicated analysis that accounts for the fields of view observed by HESS and determines a limit on the photon fraction in a given energy interval can strengthen these constraints.  Improved understanding of the underlying astrophysical electron spectrum would also allow for tighter constraints from high-latitude emission, while avoiding uncertainties associated with TeV Galactic Center emission, which is highly profile-dependent~\cite{Mack:2008wu} and would not apply here if substructure is depleted near the center of the Galaxy.

%
%--------------------------------------------------------------------%
\section{Inverse-Compton Gamma Rays}
Absent a means of containing them, high-energy electrons resulting from annihilations will escape subhalos without difficulty.  Far from the Galactic disk, the most important loss channel is inverse-Compton scattering on the CMB (we neglect the cosmic IR background, which has energy density a few percent that of the CMB~\cite{Hauser:2001xs}), since the magnetic fields there should be small~\cite{Kalberla(1998),Widrow:2002ud} and result in negligible synchrotron losses.  To calculate the gamma-ray flux, we must first find the equilibrium $e^- + e^+$ spectrum.  We start from the diffusion-loss equation for a spectrum of relativistic electrons, $n_e(E)$~\cite{Longair(1994),Strong:2007nh}
\begin{equation}
  \frac{dn_e}{dt} = \mathcal{D}(E)\,\nabla^2n_e(E) + \frac{d}{dE} \left[ b(E) n_e(E) \right] + Q(E)\,,  
\end{equation}
where the diffusion coefficient, $\mathcal{D}$, is assumed to be isotropic, $Q(E)$ is the source term, and $b(E)=b_0\,E^2$ is the radiative loss term, with $b_0\simeq 0.3 \times 10^{-16}\,$GeV$^{-1}\,$s$^{-1}$ for the CMB (in the Thomson limit).  For dark matter, equilibrium can be assumed.  In an isotropic system, there is no dependence upon $\mathcal{D}$, since particle losses are compensated for by gains.  At $\sim 1$~TeV, the electron cooling time is $\sim 10^6$~yr, so that even if electrons propagate rectilinearly, they would only travel a distance of order the virial radius of the Milky Way.  This is likely an overestimate, since their propagation should be affected by the halo magnetic field, although its structure and strength is uncertain.  Considering the length scales relevant for electrons injected by annihilation in substructure, we make the simplifying assumption that this halo magnetic field results in the IC losses occurring near the injection point (more care is needed for smooth halo signals due to the steeper gradient in particle injection with radius~\cite{Baltz:1998xv,Delahaye:2007fr}).  This reduces the problem to a continuity equation~\cite{Blumenthal:1970gc}
\begin{equation}
  -\frac{d}{dE} \left[ b_0\,E^2 n_e(E) \right] = Q(E)\,,  
\end{equation}
which can be readily solved for a given injection spectrum.  While IB signals may vary greatly between annihilation channels, essentially all models that remain viable post-Fermi lead to nearly identical equilibrium electron spectra (up to uncertainties in the astrophysical spectrum and propagation models)~\cite{Bergstrom:2009fa,Meade:2009iu}.  With generality, we consider dark matter with $m_{\rm DM}=2.35$~TeV annihilating into two $\mu^\pm$ pairs (as in \cite{Bergstrom:2009fa}).  The calculation proceeds similarly to~\cite{Yuksel:2007dr} as
\begin{equation}
  \frac{d\Phi_{\rm IC}}{dE} = \frac{1}{4\pi} {\cal J}(\psi)\, r_\odot\, \Gamma_\odot\,
                                \frac{dN_{\rm IC}}{dE}\,,
\end{equation}
where $\Gamma_\odot$ is the local annihilation rate per volume (matched to that required to agree with PAMELA/Fermi) and the resultant IC gamma-ray spectrum per annihilation, $dN_{\rm IC}/dE$, is calculated using the methods of Ref.~\cite{Blumenthal:1970gc}.  In the inner Galaxy, synchrotron and IC losses on optical/IR photon backgrounds would result in a broad range of secondary photons~\cite{Baltz:2004bb}.  We thus consider the signal resulting from annihilations occurring beyond 20~kpc from the Galactic Center, where IC on the CMB can be safely assumed to be the dominant energy-loss mechanism (based on modeling of the Galactic optical/IR photon field~\cite{Porter:2005qx}).  Using ${\cal J}(180^{\circ})$ for $r > 20$~kpc yields the solid line in Fig.~\ref{pbr}, which can be seen from the bottom panel of Fig.~\ref{Jpsi} to be nearly isotropic.  Naively including radii interior to 20~kpc would result in the dotted line.

For the scenario considered here, the IC spectrum happens to peak at a similar energy to the pionic spectrum from cosmic-ray interactions~\cite{Stecker}.  We note that this IC signal retains less angular information concerning substructure~\cite{SiegalGaskins:2008ge,Ando:2009fp,Fornasa:2009qh} than direct gamma rays (such as IB).  In comparing to isotropic gamma-ray data~\cite{Weidenspointner,Sreekumar:1997un,collaboration:2010nz}, we have made no attempt to account for other astrophysical contributions (see, e.g., \cite{Narumoto:2006qg,SiegalGaskins:2009ux} for blazars).

The velocity-dependence of the annihilation cross section in Sommerfeld models makes calculating the cosmic signal in this scenario more complicated than in the standard picture~\cite{Ullio:2002pj}.  This requires moving beyond the assumption of a constant boost (as in \cite{Profumo:2009uf,Belikov:2009cx}) due to a dependence of the velocity dispersion and hence the cross section on halo mass.  Also, the effects of baryons on dark halos vary with mass, since low-mass halos likely were never able to retain gas to form stars, and even in halos containing dwarf galaxies dark matter governs dynamics in the inner regions~\cite{Strigari:2008ib}.  Although a detailed treatment of these matters is beyond our scope, a simple estimate based on the total annihilation rate within the Milky Way halo (and scaling the amount of mass within substructure with host halo mass) suggests that this could be a factor of a few larger than the IC flux from substructure in the Milky Way halo only (as in Fig.~\ref{pbr}) with a similar spectral shape.  We note that the scenarios assumed in Refs.~\cite{Profumo:2009uf,Belikov:2009cx} are fundamentally different than considered here, since gamma-ray measurements of the inner Galaxy can strongly constrain annihilations in the smooth halo for the density profiles considered and a velocity dependence in the enhancement was not taken into account in those studies.

%--------------------------------------------------------------------%
\section{Conclusions}
If one takes the position that astrophysical resolutions of the positron excess are untenable, then to have a viable dark matter scenario requires invoking a relatively-large annihilation cross section.  One possibility for achieving this goal is a velocity-dependent cross section due to the presence of a new medium-range force resulting in a Sommerfeld enhancement.  In this study, we have examined the observational consequences of annihilation to charged particles in the context of a halo populated with dark matter substructure.

In determining the dark matter annihilation signatures of Galactic substructure, it must be kept in mind that the microphysics (annihilation) occurs within kinematically distinct subhalos so that their macroscopic distribution sets many aspects of the problem.  We have demonstrated that for a range of models, Sommerfeld-based or otherwise, annihilations in substructure, rather than in the smooth halo, are the dominant source of the locally-measured lepton flux.  These substructure-dominated models imply associated IB and IC gamma-ray emission at high latitudes at a level accessible to (and possibly already in tension with) current observations.  We have also argued that HESS TeV electron measurements can be regarded as limits on the isotropic TeV gamma rays arising from IB.  Importantly, these new prospective signals can be tested with upcoming gamma-ray observations by Fermi and air Cherenkov telescopes.

%---------------------------------------------------------------------%
\textbf{Acknowledgments.---}%
%\begin{acknowledgments}
%
We thank John Beacom, Ben Dundee, Michael Kuhlen, Troy Porter, Chris Orban, Todd Thompson, and particularly Hasan Y{\"u}ksel and Carsten Rott for useful discussions and comments.
MDK and JSG were supported by NSF CAREER Grant No. PHY-0547102 (to JB), CCAPP, and Ohio State University.
%
%\end{acknowledgments}

%\newpage

\vspace*{-0.35cm}
%---------------------------------------------------------------------%
%\textbf{References}

\end{document}